\documentclass[10pt,sigconf]{acmart}

\usepackage{hyperref}
\usepackage{enumitem}
\usepackage{listings}
\usepackage{cleveref}
\usepackage{booktabs} %
\usepackage{xspace}
\usepackage{xcolor}
\usepackage{microtype}
\usepackage{subcaption}
\usepackage{balance}
\usepackage{colortbl}
\usepackage{amsmath,amssymb}
\usepackage[linesnumbered, ruled,vlined]{algorithm2e}

\usepackage{cleveref}

\usepackage{balance}

\makeatletter
\def\@copyrightspace{\relax}
\makeatother

\DeclareMathOperator*{\argmin}{arg\,min}

\begin{document}

\newtheorem{problem}{Problem}

\newcommand{\detectlib}{\texttt{IsoDetect}\xspace}
\newcommand{\company}{\texttt{Company X}\xspace}
\newcommand{\cond}{\textrm{pred}\xspace}
\newcommand{\dataset}{data set\xspace}
\newcommand{\datasets}{data sets\xspace}
\newcommand{\spview}{\textsf{SPView}\xspace}
\newcommand{\fjview}{\textsf{FJView}\xspace}
\newcommand{\aggview}{\textsf{AggView}\xspace}
\newcommand{\hashfunc}[1]{\textsf{hash}(#1)\xspace}
\newcommand{\hashop}{\textsf{hash}\xspace}
\newcommand{\nsc}{\textsf{NormalizedSC}\xspace}
\newcommand{\rsc}{\textsf{RawSC}\xspace}

\newcommand{\avgfunc}{\ensuremath{\texttt{avg} }\xspace}
\newcommand{\maxfunc}{\ensuremath{\texttt{max} }\xspace}
\newcommand{\minfunc}{\ensuremath{\texttt{min} }\xspace}
\newcommand{\histfunc}{\ensuremath{\texttt{histogram\_numeric} }\xspace}
\newcommand{\countfunc}{\ensuremath{\texttt{count}}\xspace}
\newcommand{\sumfunc}{\ensuremath{\texttt{sum} }\xspace}
\newcommand{\varfunc}{\ensuremath{\texttt{var} }\xspace}
\newcommand{\stdfunc}{\ensuremath{\texttt{std} }\xspace}
\newcommand{\covfunc}{\ensuremath{\texttt{cov} }\xspace}
\newcommand{\corrfunc}{\ensuremath{\texttt{corr} }\xspace}
\newcommand{\medfunc}{\ensuremath{\texttt{median} }\xspace}
\newcommand{\percfunc}{\ensuremath{\texttt{percentile} }\xspace}
\newcommand{\havingfunc}{\ensuremath{\texttt{HAVING} }\xspace}
\newcommand{\selectfunc}{\ensuremath{\texttt{select} }\xspace}
\newcommand{\ratio}{\ensuremath{\rho }\xspace}

\newcommand{\insertion}{\ensuremath{\texttt{INSERT} }\xspace}
\newcommand{\update}{\ensuremath{\texttt{UPDATE} }\xspace}
\newcommand{\delete}{\ensuremath{\texttt{DELETE} }\xspace}

\newcommand{\sysfull}{AlphaClean\xspace}
\newcommand{\sys}{AlphaClean\xspace}
\newcommand{\sysnospace}{AlphaClean}

\newcommand{\tbl}[1]{\textsf{#1}\xspace}
\newcommand{\field}[1]{\textsf{#1}\xspace}
\newcommand{\cost}{\textrm{cost}\xspace}
\newcommand{\ans}{\textsf{ans}\xspace}
\newcommand{\dans}{\Delta\textsf{ans}\xspace}
\newcommand{\cqp}{correction query processing\xspace}
\newcommand{\Cqp}{Correction query processing\xspace}

\newcommand{\reminder}[1]{{{\textcolor{magenta}{\{\{\bf #1\}\}}}\xspace}}
\newcommand{\ewu}[1]{} 
\newcommand{\mps}[1]{{{\textcolor{red}{\{\{\bf meelap:\} #1\}}}\xspace}}
\newcommand{\stitle}[1]{\smallskip\noindent\textbf{#1: }}
\newcommand{\ititle}[1]{\smallskip\noindent\textit{#1: }}
\newcommand{\btitle}[1]{\smallskip\noindent\textbf{#1}}

\definecolor{light-gray}{gray}{0.95}
\definecolor{mid-gray}{gray}{0.85}
\definecolor{green}{RGB}{0,176,80}
\definecolor{darkred}{rgb}{0.7,0.25,0.25}
\definecolor{darkgreen}{rgb}{0.15,0.55,0.15}
\definecolor{darkblue}{rgb}{0.1,0.1,0.5}
\definecolor{orange}{RGB}{237,125,49}
\definecolor{blue}{RGB}{68,114,196}
\definecolor{pop}{RGB}{0,21,245}

\newcommand{\white}[1]{{\textcolor{white}{#1}\xspace}}
\newcommand{\blue}[1]{{\textcolor{blue}{{\bf #1}}\xspace}}
\newcommand{\orange}[1]{{\textcolor{orange}{{\bf #1}}\xspace}}
\newcommand{\pop}[1]{{\textcolor{pop}{{\textit{\textbf{#1}}}}\xspace}}
\newcommand{\red}[1]{\textcolor{red}{#1}}
\newcommand{\green}[1]{\textcolor{green}{#1}}
\newcommand{\gray}[1]{\textcolor{light-gray}{#1}}

\settopmatter{printacmref=false}

\pagestyle{plain}

\title{AlphaClean: Automatic Generation of \\Data Cleaning Pipelines }

\author{Sanjay Krishnan}
\affiliation{University of Chicago}
\email{ skr@cs.uchicago.edu}

\author{Eugene Wu}
\affiliation{\institution{Columbia University}}
\email{ewu@cs.columbia.edu}

\begin{abstract}
  \sloppy
The analyst effort in data cleaning is gradually shifting away from the design of hand-written scripts to building and tuning complex pipelines of automated data cleaning libraries.
Hyperparameter tuning for data cleaning is very different than hyperparmeter tuning for machine learning since the pipeline components and objective functions have structure that tuning algorithms can exploit.
This paper proposes a framework, called \sys, that rethinks parameter tuning for data cleaning pipelines.
\sys provides users with a rich library to define data quality measures with weighted sums of SQL aggregate queries.
\sys applies generate-then-search framework where each pipelined cleaning operator contributes candidate transformations to a shared pool.  Asynchronously, in separate threads, a search algorithm sequences them into cleaning pipelines that maximize the user-defined quality measures.
This architecture allows \sys to apply a number of optimizations including incremental evaluation of the quality measures and learning dynamic pruning rules to reduce the search space.
Our experiments on real and synthetic benchmarks suggest that \sys finds solutions of up-to 9x higher quality than naively applying state-of-the-art parameter tuning methods, is significantly more robust to straggling data cleaning methods and redundancy in the data cleaning library, and can incorporate state-of-the-art cleaning systems such as HoloClean as cleaning operators.
\end{abstract}

\maketitle

\section{Introduction}\label{intro}\sloppy
Data cleaning is widely recognized as a major challenge in almost all forms of data analytics~\cite{nytimes}. 
Analysts report spending upwards of 80\% of analysis time during data cleaning and preparation.
Improperly handled errors can affect the performance and accuracy of downstream applications such as reports, visualizations, and machine learning models.
In response, the research community has developed a number of sophisticated data cleaning libraries for detecting and repairing errors in large datasets~\cite{dc, rekatsinas2017holoclean, DBLP:journals/pvldb/KrishnanWWFG16, DBLP:conf/sigmod/ChuIKW16, mudgal2018deep, doan2018toward}.
As a consequence, the burden on the analyst is gradually shifting away from the design of hand-written data cleaning scripts to building and tuning pipelines of automated data cleaning libraries~\cite{krishnan2016hilda}.

Systems to automatically optimize these pipelines and their parameters are desirable.
An initial architecture is to directly apply recent hyperparameter tuning approaches for machine learning pipelines and neural network model search~\cite{li2017hyperband, sparks2017keystoneml, baylor2017tfx, golovin2017google, liaw2018tune}.
We can treat an entire data cleaning pipeline as a parametrized black box exposing tunable parameters such as confidence thresholds and editing penalties. We can quantify the success or failure of a parameter setting with a final data quality objective function (e.g., number of tuples violating integrity constraints or cross-referencing with master data).
The tuning system will then search over possible parameter settings to optimize the objective function.

Hyperparameter tuning systems are fundamentally ill-suited for data cleaning applications.
They only assume query access to the final objective value and neglect any structure and opportunities for shared computation in the pipeline.
For example, even if the objective function was based solely on integrity constraint violations, a black-box tuning system would not recognize that integrity constraints can be incrementally checked without full re-computation~\cite{fan2014incremental}.
Similarly, these systems would not recognize opportunities for re-ordering the application of libraries and excluding irrelevant libraries.

We present a new framework called \sys whose main insight is that a common intermediate representation for repairs can facilitate more efficient data cleaning pipeline optimization.
Many popular data cleaning libraries actually ``speak the same language'', where all of their repairs can be cast as cell-replacements operations~\cite{rekatsinas2017holoclean,DBLP:conf/sigmod/ChuIKW16, DBLP:journals/pvldb/KrishnanWWFG16}.
In \sys, rather than treating the entire pipeline as a single parameterized black-box, the system assess the fine-grained repairs from each parameter setting and re-orders, excludes, and merges accordingly.

Users interface their existing data cleaning libraries to \sys with minimal code that exposes an input interface to set parameters and an output interface to collect proposed edits to a table.
Libraries can be as narrow or as general as the user desires.
For example, they can be domain specific string matching functions or entire data cleaning systems such as HoloClean~\cite{rekatsinas2017holoclean}.
Each of these libraries suggests \emph{candidate repairs} to a central pool.
Users define a data quality function (the objective) with SQL aggregation queries (allowing for UDAF's) over the input table.  This subsumes popular quality measures such as integrity constraint violations~\cite{ilyas2015trends} and numerical outlier detection~\cite{bailis2017macrobase}, and can readily express application-specific quality measures such as machine learning training accuracy or goodness-of-fit to a model. 
Separate threads search through the pool of candidates to decide a sequence of repairs to construct (a cleaning pipeline) that optimizes this quality function.
\sys works in an ``anytime'' fashion where results are progressively returned to the user.

The search algorithm is implemented as a greedy tree-search that sequences the repairs~\cite{russell2016artificial}.
The space of possible repair sequences is enormous (our experiments encounter branching factors in the millions).
Thus, it is important to avoid fully evaluating a path's quality and expanding unpromising paths.    
\sys dynamically learns a model to avoid executing the pipeline and quality function in order to evaluate a given path, and can be tuned to have a low false positive rate when pruning candidate paths.  
Furthermore, the tree search can be easily parallelized across both candidate paths, as well as across partitions of the dataset based on properties of the quality function.
We use periodic synchronization to update the prediction model across parallel searches and merge transformations that repair disjoint sets of tuples. 

\sys contributes a new architecture to data cleaning optimization.
 This flexibility in composing different quality functions can help users across different domains evaluate different notions of quality within a single system. 
 The intermediate representation and generate-then-search paradigm allows for intelligent composition of multiple systems. 
Even in cases where a existing cleaning system is specifically designed for the errors in the dataset (e.g., integrity constraints), \sys can combine other cleaning operators to further improve the repairs.  Our experiments show that one of the most powerful benefits of \sys comes from the ensembling effects and its natural robustness to redundancy and/or distracting pipeline components.

\section{Background}\label{s:background}
We study parameter tuning for systems that address \emph{cell inconsistencies}, where record values are missing, incorrect, contain inconsistent references to the same entities, or contain artifacts from the extraction process. 

\ewu{Different pruning rules, samples of generated pipelines at final stage and during merging.  }

\subsection{Motivation}
Our goal is to develop techniques to automatically generate and tune data cleaning pipelines based on user-specified quality characteristics.  Thus, the user can primarily focus on composing and expressing data quality issues, and allow the system to explore the space of physical cleaning plans.  We would like the search procedure to be \textbf{progressive}, in the sense that it quickly generates acceptable cleaning plans, and refines those plans over time.  Thus, the user can immediately assess her hypothesis, or test multiple hypotheses in parallel.

\begin{figure}[t]
  \centering
 \includegraphics[width=\columnwidth]{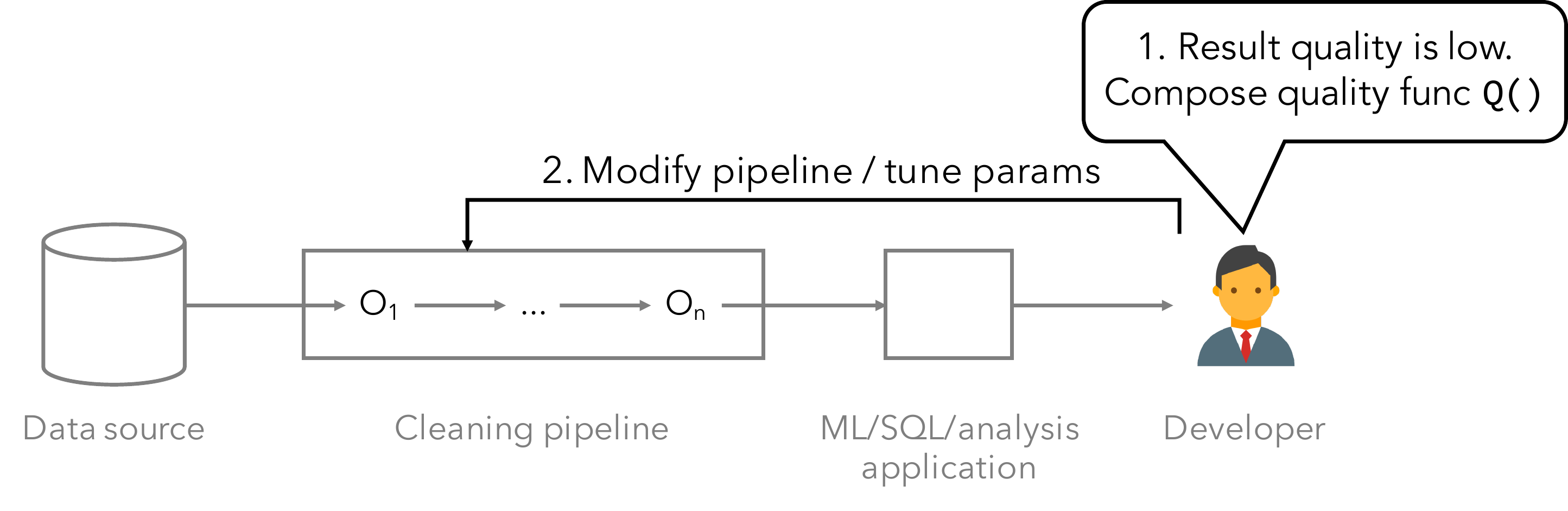}
 \caption{\small Typical data cleaning pipeline.  The user finds that analysis results (of SQL query, ML model, web application, etc) are suspicious and iteratively (1) composes a quality function to characterize the suspicious quality issues, and (2) modifies the data cleaning pipeline to address the errors.  \sys improves this human-in-the-loop process by providing an expressive, composable quality function, and automatically searching for cleaning pipelines.  \label{fig:user-pipeline}}
\end{figure}

This iterative pattern makes data cleaning a human-in-the-loop problem, where the developer explores a large space of data quality issues {\it and} data cleaning programs (Figure \ref{fig:user-pipeline}).  However, the data cleaning systems ecosystem is diffuse, with separate systems for constraint resolution ~\cite{rekatsinas2017holoclean}, cleaning in machine learning pipelines~\cite{DBLP:journals/pvldb/KrishnanWWFG16}, entity resolution~\cite{mudgal2018deep, doan2018toward}, and crowdsourcing~\cite{DBLP:journals/pvldb/HaasKWF015}.
Each of these systems has its own idiosyncrasies and parameters, and tuning even one of these systems can be a daunting challenge.
Real-world datasets have mixes of errors~\cite{krishnan2016hilda} and often require multiple systems to clean~\cite{DBLP:conf/sigmod/ChuIKW16}.
Although these systems make it easier to construct and execute a pipeline, the space of possible operator pipelines and parameterizations of each operator is exponential in the number of operators, parameters, and pipeline depth, and is infeasible for developers to manually search.

\subsection{Challenges}
We could start by considering the recent work in \emph{hyperparameter tuning} for machine learning, which identifies the optimal assignment of hyperparameters to maximize an objective function (e.g., training accuracy for ML models).
Several systems have been built to run hyperpameter and neural network model search at scale~\cite{li2017hyperband, sparks2017keystoneml, baylor2017tfx, golovin2017google, liaw2018tune}.
For single threaded search, the state-of-the-art remains to be Bayesian optimization, e.g., Python Hyperopt~\cite{bergstra2013hyperopt}.
Since Bayesian optimization is inherently sequential, for parallel and distributed settings, the community is increasingly studying randomized and grid search schemes~\cite{li2017hyperband, liaw2018tune, golovin2017google}.
For a pipeline of up to $k$ cleaning components, we can create a parameter that represents the operator type in each of the pipeline slots, along with additional operators to tune each operator in each pipeline slot.  A hyperparameter tuning algorithm will then select and assign parameter values to a sequence of operators.
Although this approach is possible, it ignores important aspects of data cleaning problems that can enable more efficient and flexible approaches.  

\stitle{Quality Function Structure} Hyperparameter tuning algorithms are also called ``black-box'' optimization algorithms because they only assume oracular access to the optimization objective (i.e., evaluate the quality of a given plan). In contrast, the objectives in data cleaning have far more structure. 
If the objective is to minimize functional dependency violations, it would be wasteful to recompute all violations after every repair. One could incrementally evaluate update the objective from the set of modified keys.
This is also true in time-series data cleaning problem where quality measures are tied to certain windows of data--there is no point re-evaluating the whole objective if only a small window is affected.
In other words, data quality measures commonly support efficient incremental evaluation, and satisfy properties that enable data partitioning.  
Neglecting this structure leads to a large amount of duplicated effort for every parameter setting evaluated.

\begin{figure}[t]
\centering
 \includegraphics[width=0.9\columnwidth]{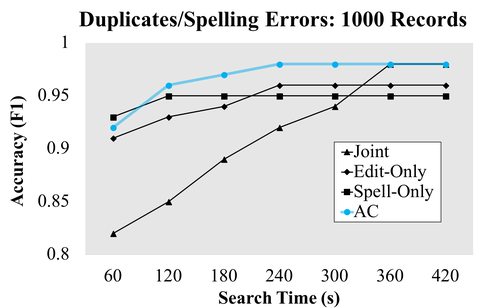}
 \caption{\small 10\% of a dataset of dictionary words are duplicated with randomly generated spelling errors. The dataset is to be cleaned with a similarity matcher and a spell checker. Holistically, tuning the parameters of both with \textsf{python hyperopt} (BB-Full) is inefficient due to interactions between the two data cleaning options. It takes over 3x the amount of search time for the joint optimization to exceed the best tuned single cleaning method (BB-Edit and BB-SpellCheck) \label{fig:teaser}}
\end{figure}

\stitle{Data Cleaning Method Structure} Similarly, black-box search algorithms would treat the data cleaning pipeline as a monolithic parametrized unit.
This leads to an attribution problems, namely, which parameter change is responsible for an increase (or decrease) in objective value.
Figure~\ref{fig:teaser} illustrates this concern on a toy data cleaning problem, with a hyperparameter search based on Tree-structured Parzen Estimator (TPE)~\cite{shahriari2016taking}\footnote{Implemented using \textsf{python hyperopt}}.  We corrupted 1000 dictionary words so that 10\% are duplicated with randomly generated spelling errors affecting 1-3 characters. The quality function is the F1 score of the cleaned dataset as compared to the ground truth.  We consider two parameterized operators: \texttt{edit\_dist\_match(thresh)} is a string edit distance similarity matcher with a tunable threshold, and \texttt{ispell(rec)} is a spell checker with a tunable recommendation parameter based on the distance between the dictionary word and the misspelled word.  The two operators partially overlap in their cleaning behavior, and we will see how it affects the search problem below.   

We compare hyperparameter search for three fixed pipelines:  single-operator pipelines (\texttt{edit\_dist\_match}) and (\texttt{ispell}), and a joint pipeline (\texttt{edit\_dist\_match}, \texttt{ispell}).  By fixing the operator pipeline, the search algorithm only needs to learn parameterizations of the operators.  Although we expect the joint pipeline to perform the best, Figure \ref{fig:teaser} shows that there is a trade-off between runtime and data quality (measured as F1 score).  It takes 3$\times$ amount of search time for the joint pipeline to exceed the best single-operator pipeline.    In contrast, the single operator pipelines quickly converge to an F1 score of $\ge95\%$.  The reason is because the two operators overlap in functionality (some duplicates can be fixed by \texttt{ispell} or \texttt{edit\_dist\_match}), which forces the join optimization to explore redundant parameter settings that have the same cleaning results.  In practice, pipelines and the set of operators can be much larger, thus the likelihood of redundant operators, or even operators that reverse changes made by previous operators, is high.

But this issue is often not present in data cleaning problems.
If we consider data cleaning operators that preserve schema (same input and output types), they can be reordered, queried/optimized independently, and ensembled in ways that general machine learning pipelines cannot.
For example, what if we independently optimized both single-operator pipelines (\texttt{edit\_dist\_match}) and (\texttt{ispell}), and then took the consensus between their repairs?
Such operations are disallowed in current hyper-parameter tuning approaches. 

This is the main intuition behind \sys: rather than treating a pipeline as a monolithic parametrized unit, we decompose it into its constituent repairs.
The system then interleaves those repairs that improve the objective function.
These repairs can be generated asynchronously in a thread of workers that query each cleaning operator with different parameters--making it robust to operators that are slow or straggle on difficult datasets.
We also include the curve for when solving this problem with \sys on Figure \ref{fig:teaser}; the next sections describe the design to accelerate such cleaning problems.

\begin{figure}[t]
\centering
 \includegraphics[width=0.7\columnwidth]{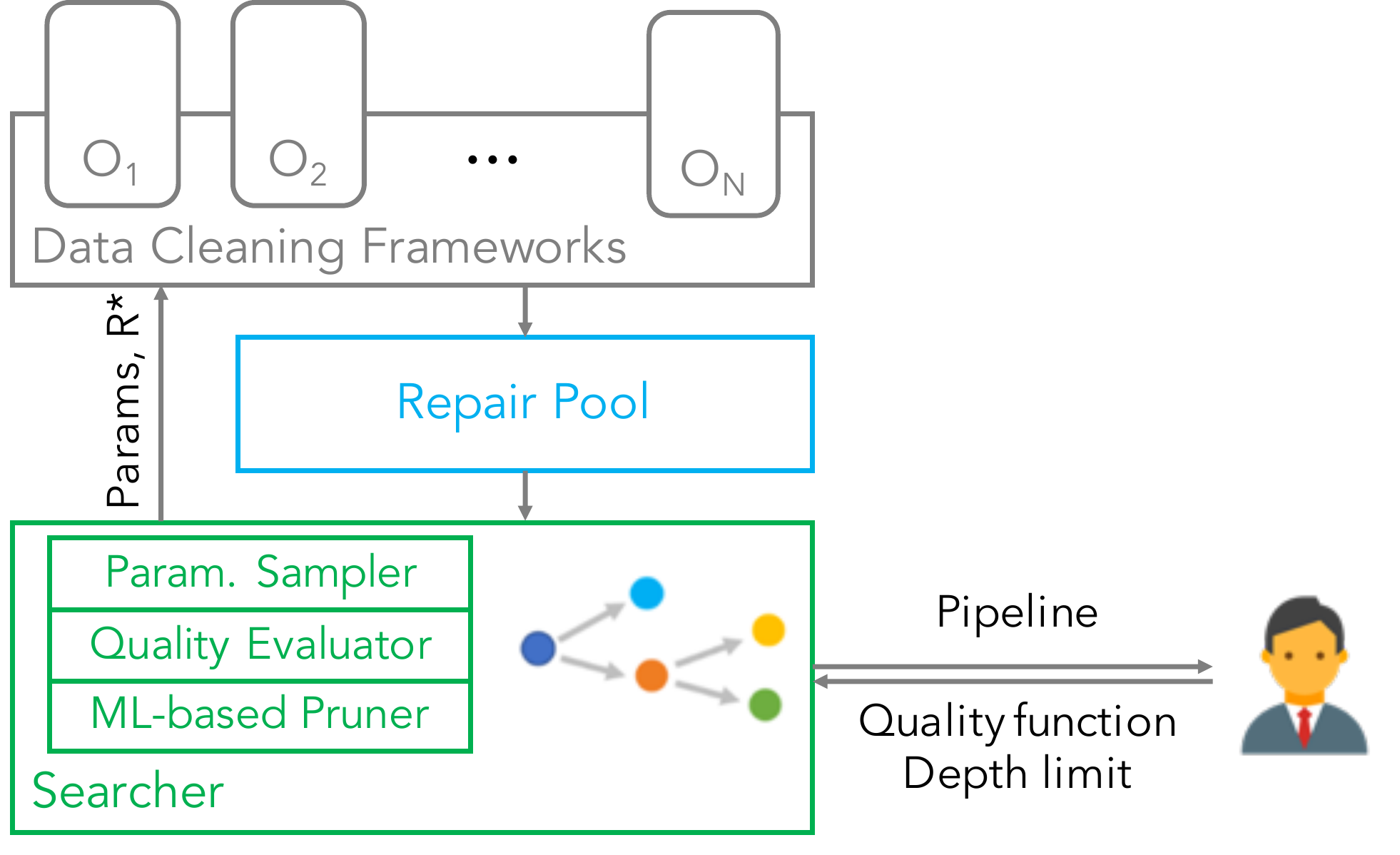}
 \caption{\small \sys decouples sampling from the parameter space from search. This allows the user to iterate quickly by observing early best-effort results. \label{fig:arch}}
\end{figure}

\section{Architecture and Overview}
Our goal is to develop a system to automatically generate and tune data cleaning pipelines based on user-specified quality characteristics.  Thus, the user can primarily focus on composing and expressing data quality issues.  We would like the search procedure to be \textbf{progressive}, in the sense that it quickly generates acceptable cleaning plans, and refines those plans over time.  Thus, the user can immediately assess her hypothesis, or test multiple hypotheses in parallel.

\subsection{Interfacing Data Cleaning Frameworks}
The first component of the system is the API interface between existing data cleaning libraries and \sys. We encapsulate the logic of such libraries into a unit we call a \emph{data cleaning framework}, a parametrized function that transforms a dataset. We assume these transformations preserve schema and do not delete/add records.
There are two classes that are important to note, \textsf{Parameter} and \textsf{Repair}.  \textsf{Parameter} is a class that represents the input parameters to a particular framework. \textsf{Repair} is a class that represents the transformations that the framework makes to a given dataset for a particular parameter setting.
\textbf{Section 4 will describe how repairs are represented and composed in more detail.}

Accordingly, each framework is then interfaced to \sys with the following API calls:

\vspace{0.5em}

\noindent Iterates through all possible parameter settings.
\begin{lstlisting}
getParameterSpace(): Iterator<Parameter>
\end{lstlisting}

\noindent Choose a particular parameter setting for the framework.
\begin{lstlisting}
setParameter(Parameter val)
\end{lstlisting}

\noindent Iterate through all repairs that framework would like to apply to the dataset.
\begin{lstlisting}
collectRepairs(): Iterator<Repair>
\end{lstlisting}

\begin{example}\it\label{e:2}
  The spell checker  \texttt{ispell(d, attr)} in Example~\ref{e:1} can be tuned by setting a maximum edit distance $d$ between the dictionary word and the attribute value $r[attr]$.  The parameter space is thus $\mathbb{N}$ for $d$, and all attributes in the relation for $attr$. Similarly, \texttt{edit\_match(d, attr)} is an edit distance matcher that searches for other \texttt{attr} values in $R$ within an edit distance $d$ of $r[attr]$, and sets $r[attr]$ to the most frequent value.  In this case, the value of the assignment is computed dynamically.  Finally \texttt{chase(fd, R)} is parameterized by a functional dependency $fd$ from a user-provided set of FDs, and will run the chase algorithm~\cite{Deutsch2008TheCR} for $fd$. 
\end{example}

\subsection{Quantifying Data Quality}
In machine learning, the objective function for hyper-parameter tuning is often given as the cross validation error of a model. In data cleaning applications, we may not always have objective ground truth. 
A quality function measures a specific notion of cleanliness for a relation, and is used as the objective function for the tuning.  
This is a \emph{proxy} for accuracy defined by the user.
These quality functions are represented in terms of SQL aggregation queries. The user provides a list of SQL aggregates (including UDAF's) and a set of weights to combine these aggregates. 
\textbf{Section 5 describes examples of quality functions and optimizations that we can apply if we have SQL descriptions.}

\begin{example}\it\label{e:3}
   In Example~\ref{e:1}, Lisa writes a functional dependency \texttt{city\_name$\rightarrow$city\_code} check as one quality function.  She writes another query that counts number of singleton city names. The final quality function is a weighted sum of the two functions.
\end{example}

\subsection{Asynchronous Architecture}
We propose a generate-then-search framework, that decouples the execution of the frameworks and pipeline quality evaluation (Figure~\ref{fig:arch}).  Each framework runs in a separate thread (or process) and continuously reruns with new parameters provided by the {\it Parameter Sampler}.   Its outputs are added to the {\it Repair Pool}.  The Searcher removes repairs from this pool to expand the set of candidate cleaning pipelines, and periodically sends the best pipelines so far to the user.   If the pool exceeds a maximum size, it applies back pressure to pause the cleaning operators until the Searcher has removed a sufficient number of conditional assignments the pool.  In practice, the cost to generate candidate assignments is far higher than the search procedure, and back pressure was not needed in our experiments.   The {\it Quality Evaluator} computes the quality of a candidate pipeline.
To make this framework practical, there are several search optimizations and heuristics that we use.
\textbf{Section 6 describes the search algorithm in detail.}

\subsection{Discussion} 
The key benefit of this asynchronous approach is that the search process does not block on a straggler cleaning framework.  It is common that parameters affect their runtime.  For example, inference thresholds and partitioning parameters can have ``cliffs'', where a small change in parameters can drastically slow down the performance of the method. Including such parameter settings in the search process naively would block the entire system.  In contrast, \sys will simply sample from faster operators until the slow inference task completes.    In fact, this design explicitly highlights the connection between the explored search space and resource scheduling.  For instance,  allocating more CPU resources to more promising operators can affect how the search space is explored.

One drawback of the asynchronous approach is that the Parameter Sampler is oblivious of the search process, so the cleaning operators may generate repairs that are not useful.   The {\it Parameter Sampler} does not attempt to preferentially sample from ``more promising'' parameter spaces, and simply uses uniform sampling.  Similarly, the Library does not perform resource scheduling, and simply allocates one thread per cleaning operator, and each process executes parameter assignments serially. 
We will show that using machine learning to identify promising search paths can alleviate this concern.

\section{Repairs}\label{s:problem}
A main insight of \sys is that a common intermediate representation for repairs can facilitate more efficient data cleaning pipeline optimization.
Data cleaning frameworks that are interfaced to \sys asynchronously pool together suggested data repairs.

\subsection{Repair Format}
Repairs are specified as ``conditional assignments'', which are sentences of the form ``if a tuple satisfies a condition, then set an attribute to a specified value'': 
{\small\begin{verbatim}
    ca(r):
      if pred(r): r[attr] = v
      return r
\end{verbatim}
}
The predicates are in a restricted language consisting of equality clauses and single attribute inequalities, e.g., 
{\small\begin{verbatim}
r[city_code] == 'NY'     
r[id] > 3      
r[name, code] == ('New York', 'NY')
\end{verbatim}
}
This restriction on predicates allows for efficient conflict testing; determining whether two repairs are independent of each other. 
Despite the restriction, it is still expressive enough to capture many important types of data cleaning. Because, in the degenerate case, we can simply use the tuple's primary key as a a predicate attribute. In that case, for each tuple there is a separate conditional assignment. 

\begin{example}\it
  \texttt{ca(code.prefix(``NY''), code, ``NYC'')} sets the code to ``NYC'' for all records where the city code starts with ``NY''.  This single condition could be replaced with three operations with predicates  \texttt{id=2}, \texttt{id=3}, \texttt{id=4} {\it for the example table}, where operation can be executed and added to a cleaning pipeline independently. 
 \end{example}
 
 The interesting cases are when we can aggregate repairs together under a single, more-informative predicate. For example, we often find this with numerical outlier detection libraries that identify a threshold on attribute values after which they are deemed outliers.
 
 \begin{example}\it
  \texttt{ca( Population < 0, Population, NULL')} sets a population attribute to NULL for all records where it is less than 0. 
 \end{example}

\subsection{Operations Over Repairs}
A cleaning pipeline is defined as composition of conditional assignments, where $(ca_2 \circ ca_1)(r) = ca_2(ca_1(r))$. Note that $ca_1$'s changes may be overwritten by $ca_2$.  A composition can similarly be evaluated over a relation $R$: $(ca_2\circ ca_1)(R) = ca_2(ca_1(R))$.    
The next section will describe the interface to evaluate the quality of a cleaning pipeline in more detail. The basic search problem that underpins \sys is a search over possible compositions of conditional assignments to optimize a quality function $Q$:
\begin{problem}[Search Problem]%
Given quality function $Q$, a set of frameworks $L$, relation $R_{dirty}$, find valid plan (composition of conditional assignments) $p^* \in \mathcal{P}$ that optimizes $Q$:
\[
p^* = ~ \argmin_{p \in P} Q( p(R_{dirty}) ).  
\]
\end{problem}
$p^*(R_{dirty})$ returns the cleaned table, and $p^*$ can potentially be applied to any table that is union compatible with $R_{dirty}$. 

In general, conditional assignments are not commutative, meaning that $c_i\circ c_j \ne c_j\circ c_i$.  However, the intermediate representation allows us to efficiently test if conditional assignments are commutative and can be run in parallel.
If $c_i$ and $c_j$'s predicates are non-overlapping and their assignment values do not affect this independence, then their operations are commutative ($c_i\circ c_j = c_j\circ c_i$). 
Based on this observation, we use a heuristic to opportunistically merge candidate pipelines if they are disjoint in such as way and both pipelines independently increase the quality function.

\section{Quality Functions}
Now, we have to define the API for assessing the quality of a pipeline.
A quality function measures a specific notion of cleanliness for a relation, and is used as the cost model for the pipeline search.  
Our goal is to define these quality functions in a sufficiently ``white box'' way to be able share computation when possible over different search expansions.

\subsection{SQL Aggregate Queries}
Quality functions are defined is in terms of SQL aggregation queries. For example, the number of functional dependency violations (e.g., $\texttt{city\_name} \rightarrow \texttt{city\_code}$) is expressible as:
{\small\begin{lstlisting}
  q1(T): SELECT count(1)
         FROM T as c1, T as c2,
         WHERE (c1.city_name == c2.city_name) AND
               (c1.city_code <> c2.city_code)
\end{lstlisting}}
\noindent Conditional functional dependency violations is a well-studied quality function, and many systems optimize for this class of objectives~\cite{rekatsinas2017holoclean,DBLP:conf/sigmod/ChuIKW16}.   

However, this example highlights that \emph{even seemingly simple data cleaning problems can require the flexibility to express multiple quality functions.}   For example, record 1 does not violate the above functional dependency, and will be missed by most functional dependency solvers.  Suppose the analyst observed a histogram of city names and noted that there were a large number of singleton entries. Thus, she could write a second quality function that counts the number of singleton entries.  This is an example of a quality measure that other systems such as Holoclean and Holistic Data Cleaning do not support as input~\cite{rekatsinas2017holoclean,DBLP:conf/sigmod/ChuIKW16}:
{\small
\begin{lstlisting}
  q2(T): SELECT count(1)
         FROM ( SELECT count(1) as cnt FROM T,
                GROUP BY city_name HAVING cnt = 1)
\end{lstlisting}}
Finally, the user can embed the downstream application as a user defined function (UDF).  For instance, the machine learning model accuracy can be added as a quality function that calls a UDF \texttt{model.eval()}.  In our experiments using the London Air Quality benchmark, we show how a parametric auto-regressive model that measures curve smoothness can be expressed as a quality function:
{\small\begin{lstlisting}
  q3(T): SELECT avg(err) AS acc
         FROM ( SELECT model.eval(X) = Y FROM T )
\end{lstlisting}}
\noindent \sys lets the user compose linear combinations of quality functions together. We model the composition over $n$ individual quality functions as $Q(T) = \sum_{i=1}^n w_iq_i(T)$.  For example, $n=2$ in the example, and captures the semantic functional dependency issues as well as the syntactic string splitting errors in a single cost model.  Our experiments simply set $w_i=\frac{1}{n}$.

We designed the quality function in this way for several reasons.  SQL aggregations can be incrementally computed and maintained, and can be efficiently approximated.  This is important because each conditional assignment typically modifies a small set of records, and thus allows efficient re-computation that scales to the number of cleaned records rather than the size of the dataset.  The linear compositions enables parallelization across each $q_i$ term, and the aggregation functions are typically algebraic functions that can be parallelized across data partitions.  The combination of incremental maintenance, and data and quality function parallelization speeds up evaluation by up to 20x in our experiments.

\subsection{Incremental Maintenance}\label{s:qualityivm}
Most cleaning operators modify significantly fewer records than the entire dataset.  Since quality functions are simply aggregation queries, \sys can incrementally evaluate the quality function over the fixed records rather than the full dataset.  
This is exactly the process of incremental view maintenance, and we use standard techniques to incrementally compute quality functions.

Suppose we have relation $R$, quality function $q$, and a set of conditional assignment expressions $C$.  When possible, \sys computes \red{$q(R)$} once and then for each of the expressions $c \in C$ compute a delta such $q(c(R)) = \red{q(R)} + \delta_c(q(R))$.
For many types of quality functions such incremental computation can be automatically synthesized and can greatly save on computation time.
Currently, this process is not automatic and \sys relies on programmer annotations for incremental updates.
It is not hard to automate this process when possible, but this is orthogonal to the topic studied in this paper.
The property that we would have to test is self-maintanability~\cite{gupta1996data}, and we would have to implement delta computation in relational algebra. 

Let us consider a concrete example with the quality function $q_1$, a functional dependency checker, from the previous section.
$R'$ is the resulting relation after applying \texttt{c} to all of the records.
Let $r_{pred}$ be the set of records that satisfy the predicate of the conditional assignment expression and $r_{pred}'$ be the resulting transformed records.
$q_1$ can be expressed in relational algebra in the following way:
\[
q_1(R') = \textsf{count}( R' \bowtie R' )
\]
$R'$ can be described in terms of $R$:
\[
R' = R - r_{pred} + r_{pred}' 
\]
leading to the following expression:
\[
q_1(R') = \red{q_1(R)} - \textsf{count}( r_{pred} \bowtie R )  + \textsf{count}( r'_{pred} \bowtie R )
\]
Evaluating this quality function using a hash join reduces the incremental evaluation cost to roughly linear in the size of the number records changed, rather than the size of the relation.

\section{Search Algorithm}\label{s:search}
This section describes our system optimizations.

\begin{figure}[t]
    \centering
    \includegraphics[width=\columnwidth]{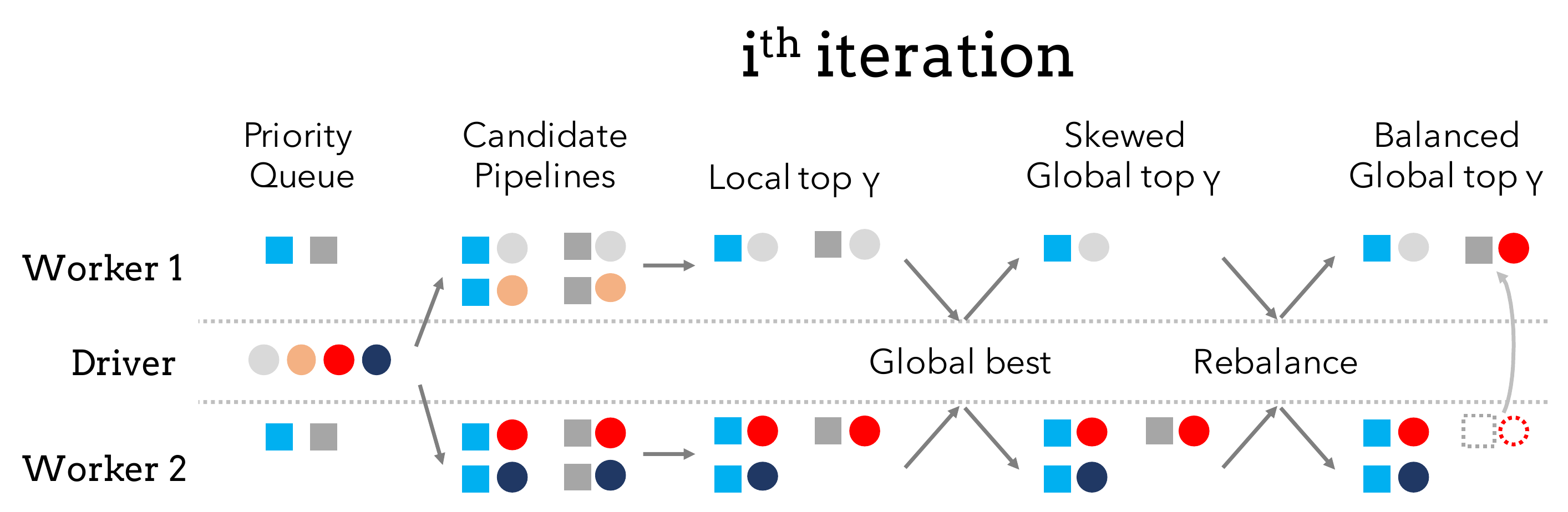}
    \caption{In each iteration, each worker starts with a subset of the priority queue (boxes).  The driver distributes conditional assignments (circles) to generate candidate pipelines (box-circles).  A series of synchronization points identify the globally top $\gamma$ candidates and redistributes them across the workers.   \label{fig:algo}}
\end{figure}

{
\begin{algorithm}[t]
\KwData{Q, S}

 Pruned = $\{\}$  //empty priority queue

 \For{$s \in S$ }{
 
    $\bar{s}$ = $s$
    
    \For{$s' \in S$ }{
        
        $m$ :=  $s' \circ \bar{s}$
        
        $\bar{s}_{pred}$ := $\cup_{c\in \bar{s}} c.pred$
        
        $s_{pred}'$ := $\cup_{c\in s'} c.pred$

        \If{$\bar{s}_{pred} \cap s_{pred}' = \emptyset$ and $Q(\bar{s}) < Q(m)$}{
        
           $\bar{s} = m$
        }
        
    }
    
    Pruned.push($\bar{s}$)
        
}

\Return Pruned
\caption{Pruning Disjoint Paths}
\label{alg:pruning}
\end{algorithm}
}

\subsection{Parameter Sampler}
By default, users simply specify an operator's parameter domain as a list of values, and the Parameter Sampler uniformly samples from the domain. Non-uniform sampling is possible, and worth exploring in future work.   In addition, users can specify two types of parameter properties, for which \sys can apply search optimizations:

\begin{itemize}[leftmargin=*,topsep=.3em,itemsep=-.2em,partopsep=-.5em]
  \item \stitle{Attribute Name Parameters} If the parameter represents an attribute in the database, then \sys can infer the domain of allowable values.   For example, a numerical outlier detection algorithm might apply to a single attribute or a subset of attributes. \sys can also prune the paramater space by pruning attribute names that are irrelevant to the quality function.  

  \item \stitle{Threshold Parameters} Numeric parameters are often used as thresholds, inference parameters, or confidence bounds.  For these, users specify the most and least restrictive ends of the value domain, and \sys will sweep the space from most to least restrictive.   For instance, \texttt{ispell} only uses the dictionary if the attribute value is within \texttt{rec} characters of the dictionary word.  Thus, \sys will initially sample $\texttt{rec}=0$ and gradually relax the threshold.   
\end{itemize}

\subsection{Parallelization} 
Even with incremental evaluation, composing and evaluating $Q(s'(R))$ is the single most expensive search operation.  Thus, we parallelize across candidate pipelines and data partitions. The prototype uses Ray~\cite{ray} to schedule and parallelize  over multiple CPUs and machines.

\stitle{Search Parallelism}
Conceptually, we execute all expansions for a given plan $s$ in parallel.  We materialize the incremental deltas in memory, and evaluate the quality of each $s' = c\circ s$ in parallel using a thread pool.  Each thread drops a given $s'$ if its quality is lower than $\gamma\times$ the maximum quality from the previous \texttt{WHILE} iteration or the local thread.  At the end of the \texttt{WHILE} iteration, the threads synchronize to compute the highest quality, and flush the remaining candidates using the up-to-date quality value.

The implementation of this conceptual parallelization is a little bit more complex. 
Each worker is given a subset of candidate pipelines to locally evaluate and prune, and the main challenge is to reduce task skew through periodic rebalancing.  We use a worker-driver model with $j$ workers (\Cref{fig:algo}).

Let $S^{next} = S\times P$ be the set of candidate pipelines (e.g., \includegraphics[height=8pt]{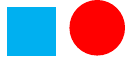}) to evaluate in the current iteration of the search algorithm. For instance, $S=\{NOOP\}$ in the first iteration, so the candidates are the set of individual data transformations $P$.   The driver assigns the input relation $R$ and $\frac{1}{j}$ of $P^{next}$ to each worker.   Each worker evaluates and computes the top-$\gamma$ candidates based on the best worker-local quality.   The worker runs and caches the parents of its assigned candidate pipelines (\includegraphics[height=8pt]{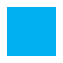}, \includegraphics[height=8pt]{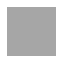}) to incrementally compute the quality function.
  
Note that the worker-local top-$\gamma$ candidates are a superset of the top-$\gamma$ global candidates because the best local quality is $\le$ the global best.   Thus the workers synchronize with the driver to identify the global best candidate and further prune each worker's top candidates.  At this point, all candidate pipelines are within $\gamma$ of the globally best candidate, but their distribution across the workers can be highly skewed.  \sys performs a final rebalancing step, where each worker sends the number of un-pruned candidates to the driver.  Workers with more than $\frac{1}{j}$ of the total number redistribute the extras to workers with too few candidates.  When redistributing, workers communicate directly and do not involve the driver (e.g., Worker 2 sends \includegraphics[height=8pt]{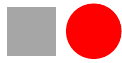} to Worker 1).   If the total number is $<k$, then candidates are randomly chosen to be replicated.  Only the pipelines and their qualities are sent; the pipeline results are re-computed by the receiving worker.  This ensures that the priority queue in the next iteration is evenly distributed across all workers.

\stitle{Data Parallelism}
Many large datasets are naturally partitioned, e.g., by timestamp or region. 
The idea is to partition the dataset in such a way that errors are local to a small number of records.
This means that a fix for a given record does not affect the other records outside of the partition.
There is a relationship between partitioning and the quality functions defined.
For example, quality functions derived from functional dependencies can define blocks by examining the violating tuples linked through the dependency.  Similarly, users can define custom partitioning functions.  In our current implementation, we partition the input relation by row by user-specified blocking rules.

\subsection{Learning Pruning Rules}\label{s:pruning}

Traditionally, data cleaning systems manually implement pruning heuristics for a fixed quality function that can work for any dataset. For example, the Chase~\cite{} used in functional dependency resolution does not make an edit to the table unless it enforces at least one tuple's FD relationship.  Similarly, in entity matching problems, one restricts the search to only tuples that satisfy a blocking (clustering) criteria.  These can be viewed as pre-conditions over the search space.

Our idea is to {\it learn} pruning rules that are data and quality function dependent.  Our hypothesis is that data errors are often systematic in nature, and correlated with specific attributes and their values.  Our pruning optimization seeks to distinguish conditional assignments that are likely to contribute to the final cleaning pipeline, and those that will not.  To do so, the basic strategy is to independently execute the Search algorithm on partitions of the dataset, and learn a prediction model.

\stitle{Approach}
As described in Section~\ref{s:search}, \sys uses data parallelism to execute search for each block of the dataset in parallel.  Thus, each block results in an optimal cleaning pipeline.  \sys models the optimal cleaning pipeline for each block as a set of training examples.  Specifically, cach conditional assignment $c$ in a block's optimal cleaning plan $s*$ can be labeled as a positive training example, while all other conditional assignments that were not used are negative examples.  

As \sys processes more blocks, it trains a classifier to predict whether a given transformation will be included in the optimal program, based on the training examples across the blocks.  In our approach, the prediction model $M(p): P \mapsto \{0,1\}$ is over the data transformations and not the data; is this sense, \sys learns pruning rules in a dynamic fashion. 
New expansions are tested against the classifier before the algorithm proceeds.
Internally, \sys uses a Logistic Regression classifier that is biased towards false positives (i.e., keeping a bad search branch) over false negatives (e.g., pruning a good branch). This is done by training the model and shifting the prediction threshold until there are no False Negatives. 

\stitle{Featurization}
Note to use this approach, we have to featurize each conditional assignment $c$ into a feature vector.
We \emph{do not} featurize the data as in other learning-based data cleaning systems.
Now, we describe how each conditional assignment is described as a feature vector.
Let $A$ be a list of all of the attributes in the table in some ordering (e.g., $[city\_name, city\_code]$).
Every conditional assignment statement is described with a predicate \texttt{pred}, \texttt{targ} a target attribute, and a target \texttt{value}. 
$A_{pred}$ is the subset of attributes that satisfy the predicate and $A_{target}$ is the singleton set representing the target attribute.
Each of these sets can be turned into a $|A|$-dimensional binary vector, where 1 represents presence of an attribute, and we call these vector $f_{pred}$ and $f_{target}$ respectively.
Then, we include information about the provenance of the conditional assignment $c$, from which data cleaning method it was generated and what parameter settings.
This feature called $f_{dc}$ is contains a 1-hot feature vector describing which is the source data cleaning method and any numerical parameters from the source method. 

We believe this is one of the reasons why a simple best-first search strategy can be effective.  For the initial blocks, \sys searches without a learned pruning rule in order to gather evidence.  Over time, the classifier can identify systematic patterns that are unlikely to lead to the final cleaning program, and explore the rest of the space.  
The features guide \sys towards those data cleaning methods/parameter settings that are most promising on previous blocks.
 \sys uses a linear classifier because it can be trained quickly with few examples.   However, we speculate that across a sufficient number of cleaning problems that share common set of data transformations (say, within the same department), we may adapt a deep learning approach to automatically learn the features themselves.

\begin{figure*}[ht]
\centering
 \includegraphics[width=0.32\textwidth]{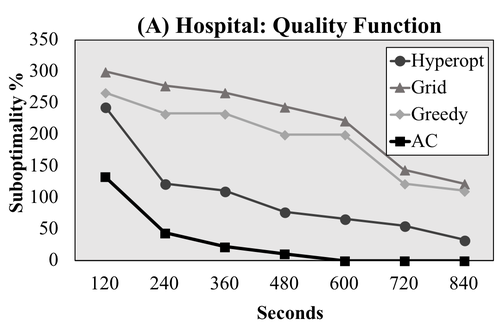}
 \includegraphics[width=0.32\textwidth]{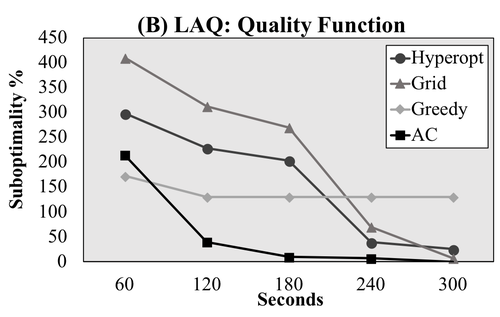}
 \includegraphics[width=0.32\textwidth]{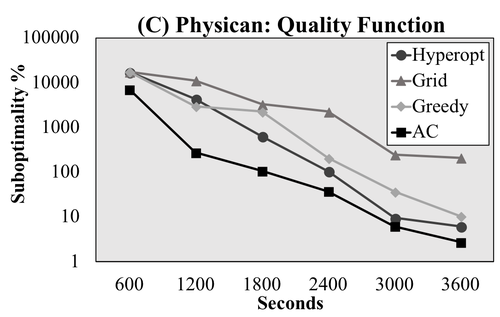}
 \caption{ Tuning Against Quality Functions. On the x-axis is the search time in seconds, and on the y-axis is the suboptimality w.r.t the quality of the ground truth data. (A) Hospital dataset, (B) London Air Quality Dataset, (C) Physician Dataset. In all three datasets, \sys converges to a more accurate solution faster than the alternatives. \label{exp2}}
\end{figure*}

\begin{figure*}[ht]
\centering
 \includegraphics[width=0.32\textwidth]{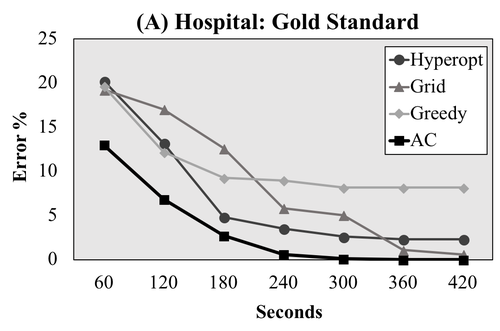}
 \includegraphics[width=0.32\textwidth]{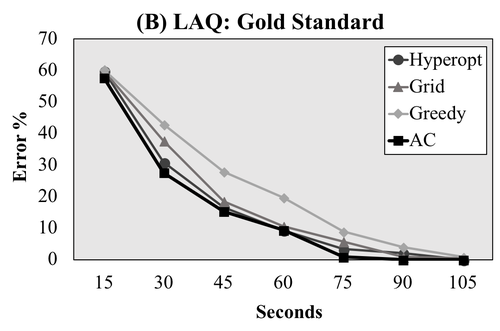}
 \includegraphics[width=0.32\textwidth]{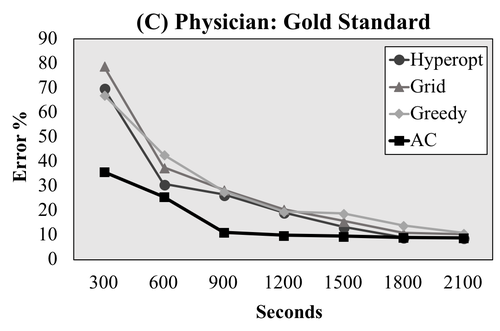}
 \caption{ Tuning Against Gold-Standard Data. On the x-axis is the search time in seconds, and on the y-axis is the inaccuracy w.r.t ground truth. (A) Hospital dataset, (B) London Air Quality Dataset, (C) Physician Dataset. In all three datasets, \sys converges to a more accurate solution faster than the alternatives.   \label{exp1}}
\end{figure*}

\section{Experiments}\label{s:exp}
Our goal is to 1) compare \sys with modern blackbox hyper-parameter tuning algorithms, 2) understand its strengths and failure cases, and 3) highlight the promise of a general search-based method through comparisons with data cleaning systems (e.g., HoloClean~\cite{rekatsinas2017holoclean}) that are specialized to specific classes of data errors.

\subsection{Datasets and Baselines}
We focus on three datasets used in prior data cleaning benchmarks.  Each dataset exhibits different dataset sizes and data cleaning needs. Each dataset also provides ground truth cleaned versions. We also describe the default cleaning operator libraries for each dataset, informed on prior benchmarks, as well as baseline hyper-parameter tuning methods.

\subsubsection{Datasets and Cleaning Benchmarks}

\stitle{Hospital} This dataset contains UK Hospital information, and used in~\cite{he2016interactive, rekatsinas2017holoclean}.  Roughly 5\% of the cells are corrupted with mispellings, missing values, or other inconsistencies.  The default quality function tries to minimize the number of singleton cities with only one hospital, because it may be due to data errors (example in Section~\ref{s:problem}).  The default library contains: \texttt{ispell.replace(thresh, attr)} as described in Section~\ref{s:problem} replaces the attribute value if it is within a threshold of a dictionary value, \texttt{minhash.replace(thresh, attr)} runs the minhash de-duplication algorithm~\cite{broder2000min} to find similar values and sets them to be equal, and \texttt{fd.replace(fd)} enforces a functional dependency with the chase algorithm~\cite{aho1979theory}.

\stitle{London Air Quality (LAQ)} The dataset contains measurements of air pollution particulate matter from London boroughs~\cite{londonair}. Around 2\% of the measurements (cells) are corrupted by a variety of different outliers including very large values as well as clipped very small values.  As the errors are mostly numerical in this dataset, the default quality function fits an autoregressive model to the windows and computes the average error to the fit model:
\begin{verbatim}
    SELECT AVG(autoregression.error(window))
    FROM data [Range 5 hours];
\end{verbatim}
\noindent The default library consists of parametrized outlier detector methods from dBoost~\cite{mariet2016outlier} and pyod~\cite{pyod}.  Both detect outliers and set them to the last known non-outlier value. \texttt{dBoost.histogram(peak\_theshold, outlier\_threshold, window\_size)} detects peaks in histograms of sliding windows of the data, \texttt{dBoost.gaussian(K, window\_size)} thresholds values outside $K$ standard deviations  from the mean of a sliding window, and \texttt{pyod.pca(outlier\_threshold, window\_size)}  applies PCA to sliding windows and thresholds them by the sum of weighted projected distances to the eigenvector hyperplane.

\stitle{Physician} The Physician Compare dataset was used in HoloClean~\cite{rekatsinas2017holoclean}, and contains information on medical professionals and the primary care practice they are associated with.  It contains misspellings, inconsistencies, and missing data. The default quality function is the set of 8 functional dependencies defined in~\cite{rekatsinas2017holoclean}.  The default library contains the operators for the Hospital dataset as well as HoloClean, which is wrapped as the operator (\texttt{holoclean.replace(fd, threshold)}) that enforces a functional dependency using HoloClean's suggested cell value fix if its confidence exceeds a threshold.

\subsubsection{Baselines}
We consider the following baseline techniques by encoding the data cleaning problem into a large set of parameters as described in Section~\ref{s:background}. To speed up the methods, we use the incremental computation optimization for quality function evaluation. Every component is given the same time limit and has to return its best cleaning strategy by that time. 

\stitle{Grid Search} We cascade all of the operators into a fixed order and treat it as a monolithic parametrized unit. We search over all possible values of the discrete parameters and a grid of values over the continuous parameters, and evalute the quality at the end. We use grid search as a baseline as it is easy to parallelize and compare at scale.

\stitle{Hyperopt} We use exactly the same setup as Grid Search but instead of searching over a grid,  we use python hyperopt to perform a Bayesian optimization and intelligently select parameters. We use hyperopt as a baseline for an optimized single-threaded search through the parameter space.

\stitle{Greedy} We tune each data cleaning algorithm independently with respect to the original data. We use a grid search scheme on each component independently. We use greedy as a baseline to illustrate the benefits and drawbacks of individual optimization of each data cleaning system.

\subsection{End-to-End Experiments}
We first compare \sys with the baseline parameter tuning methods on the three benchmark datasets.   For all three benchmarks, we add a component to the quality function that penalizes size of the changes in the cleaned dataset, based on cell-wise edit distance for string values or absolute difference for numeric values.    To understand the  relative convergence of the methods, we report suboptimality, defined as the ratio of the quality score evaluated on the ground truth over the current best quality score.  To understand the absolute cleaning improvements, we report Error, as defined by F1 of current cleaned cells with respect to the ground truth cleaned cells.

Figure \ref{exp2} plots suboptimality convergence over search time in seconds.
All searches are run with one search thread (\sys uses one extra thread to generate conditional assignments across all cleaning operators in a loop).  
We find that \sys quickly finds strong solutions quickly, because the asynchronous design quickly allows for partial data cleaning even if early parameter choices are suboptimal.
Throughout the search process, \sys is up to 9x higher quality than the next best baseline, and ultimately converges to higher quality solutions.

Figure~\ref{exp1} plots the error rate over search time, but the quality function computes the number of cells that differ from the ground truth dataset.  We consider this the best-case gold standard quality function.
We see that in this case, \sys converges to the ground truth more quickly than the next best baseline (up to $3\times$).

\subsection{Optimization Contributions}

\begin{figure}[t]
\centering
 \includegraphics[width=0.8\columnwidth]{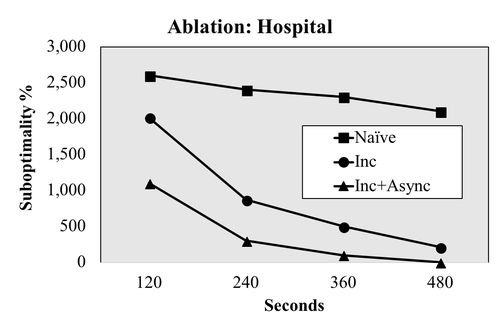}
 \includegraphics[width=0.8\columnwidth]{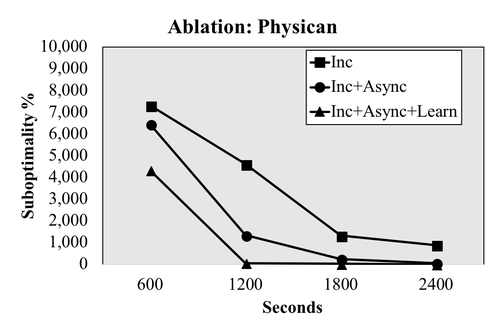}
 \caption{Contribution of the different optimizations. Incremental quality evaluation (Inc), asynchronous search (+Async), and learned pruning models (+Learning) all contribute to improved convergence above Naive. The hospital dataset is too small for learning, and the physician dataset is too large to finish without incremental evaluation. \label{exp7}}
\end{figure}

Figure~\ref{exp7} incrementally removes components of \sys to understand where the benefits come from:  incremental quality evaluation (Inc), asynchronous conditional assignment generation (Async), learning a pruning model (Learn).  We find that \sys without any optimizations (Naive) does not finish on the Physician within an hour due to the large size of the dataset, and that pruning is ineffective for Hospital due to its small size, so do not include them in the plots.

We find that all techniques are crucial.  \sys is designed to quickly evaluate a large number of quality functions, thus Inc is a primary performance optimization. 
Async allows search to quickly explore more pipelines without being blocked by conditional assignment generation, while Learn is able to effectively prune large subsets of the search space when the dataset is large; if the dataset is small there can be too few partitions from which to collect training samples.
These optimizations can improve convergence by more than 20x.

\begin{figure}[t]
\centering
 \includegraphics[width=0.8\columnwidth]{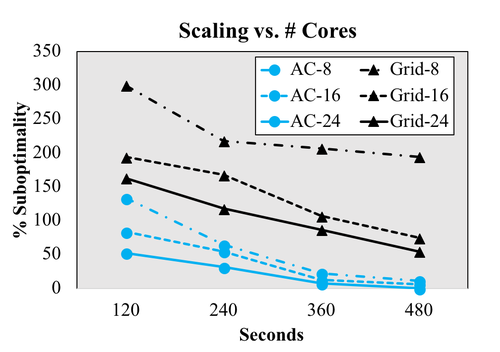}
 \caption{Scaling performance on the hospital dataset. \sys can benefit from parallelism. \label{exp3}}
\end{figure}

\subsection{\sys Performance Sensitivity}
We now study settings that affect \sys convergence rates.

\begin{figure}[t]
\centering
 \includegraphics[width=0.8\columnwidth]{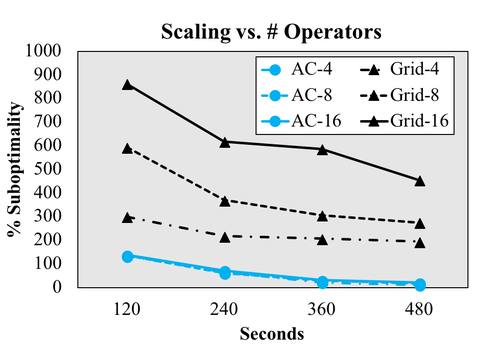}
 \includegraphics[width=0.8\columnwidth]{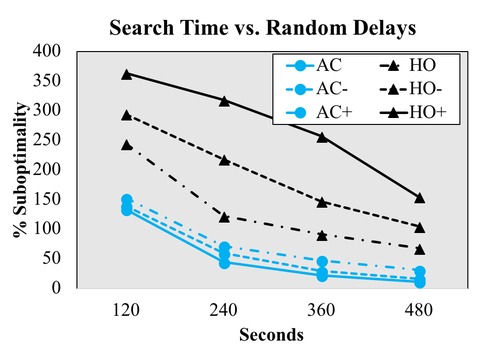}
 \caption{Both experiments are on the hospital dataset. (A) Convergence with $4, 8, 16\times$ redundant cleaning operators. (B) Convergence for short $[0-50ms]$ (AC-/HO-) and long $[0-100ms]$ (AC+/HO+) operator delays. \label{exp45}}
\end{figure}

\stitle{Scaling to Cores}
The asynchronous search architecture has desirable scaling properties.
We compare to Grid search and vary the number of threads given to both frameworks.
In \sys, we allocate one thread to each data cleaning method to generate candidate conditional assignments and the remainder to the search algorithm.
Figure \ref{exp3} illustrates the scaling on the hospital dataset.
Results suggest that \sys can benefit from parallelism.

Note that most blackbox approaches such as Grid search can run cleaning operators in parallel, however they block until the operators finish before performing a search step (picking and trying a candidate pipeline) and choosing the next parameters to try.  Thus, they can be blocked by straggler operators.  More sophisticated hyper-parameter tuning algorithms, such as hyperopt are inherently sequential and do not run cleaning operators in parallel.

\stitle{Library Size} Figure \ref{exp45}b uses the Hospital benchmark and varies the number of redundant cleaning operators, by duplicating the library by $4, 8, 16\times$.  Each duplicate runs in a separate thread.  To exploit parallelism, we compare with grid search (Grid) using the same number of threads.  \sys performs nearly identically irrespective of the redundancy, while grid search degrades considerably due to the reasons described in the above scaling experiment.

\stitle{Slow Cleaning Operators}
We use the Hospital benchmark to study robustness against slow cleaning operators. Figure \ref{exp45}a compares \sys (AC) to hyperopt (HO) with when adding random delays to the cleaning operators.  Each operator in \sys runs in a separate thread, whereas hyperopt is a sequential algorithm.  \sys is significantly more robust to these delays that hyperopt.

\begin{figure}[h]
\centering
 \includegraphics[width=0.8\columnwidth]{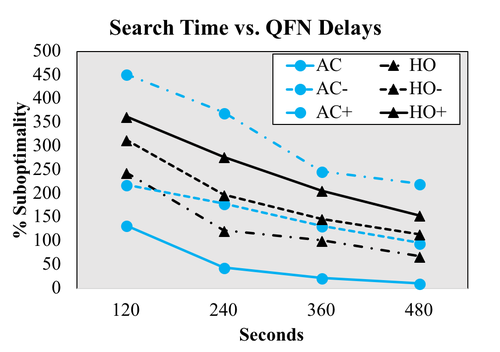}
 \caption{Convergence for short $[0-50ms]$ (AC-/HO-) and long $[0-100ms]$ (AC+/HO+) quality function evaluation delays.  \label{exp6}}
\end{figure}

\stitle{Slow Quality Evaluation}
\sys makes a design assumption that the cleaning is the bottleneck and not the quality evaluation. Figure \ref{exp6} runs the Hospital benchmark with varying delays in quality evaluation: AC-/HO- for random $[0-50ms]$ delays, and AC+/HO+ for random $[0-100ms]$ delays. While both \sys (AC) and hyperopt (HO) are affected, \sys is much more sensitive because \sys evaluates quality functions at a far higher rate than hyperopt.

\begin{figure}[h]
\centering
 \includegraphics[width=.8\columnwidth]{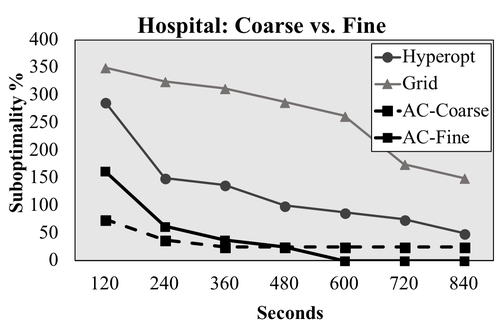}
 \caption{We apply \sys to the hospital dataset with a coarsed candidate generation scheme. Each data cleaning method produces one full-table transformation per parameter setting (AC-Coarse). While it does not converge to the global solution that the original method does (AC-Fine), it still provides a benefit due to operator exclusion and re-ordering. \label{exp8}}
\end{figure}

\stitle{Coarse vs. Fine Predicates}
Cleaning operators set the predicate granularity of the conditional assignments that they output.  Figure~\ref{exp8} evaluates the trade-off between coarse (AC-Coarse) and fine-grained (AC-Fine) conditional assignment predicates on \sys We generate coarse predicates by merging all conditional assignments generated by an operator into a single ``meta assignment'' that applies the set internally.  The main difference is that \sys cannot pick and choose from within the set.   We see that coarse predicates is initially better because \sys searches through a smaller conditional assignment pool for acceptable pipelines.  However, AC-Fine converges to a better plan because it is capable of making finer-grained decisions later on.  This suggests a potential coarse-then-fine hybrid approach in future work.  We include hyperopt and grid as reference.

\begin{figure}[h]
\centering
 \includegraphics[width=.8\columnwidth]{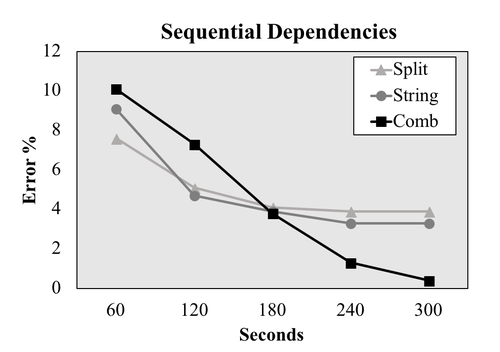}
 \caption{On a synthetic dataset with extraction and spelling errors, \sys is able to combine two types of cleaning operators (Split, String) in the appropriate sequence to clean the dataset.\label{exp10}}
\end{figure}

\stitle{Sequential Data Cleaning}
It is possible that a best-first search through an asynchronously generated candidate pool may affect problems where the precise sequence of data cleaning operations matters.
In the last experiment, we consider a synthetic dataset similar to the \texttt{City} table in Section~\ref{s:problem}.  
We construct a dataset of 10000 tuples with string attributes \texttt{str1, str2}, and functional dependency \texttt{str1$\rightarrow$str2}.  We pick 5\% of the tuples and add random spelling errors or randomly swapped values. For 50\% of that subset of tuples, we concatenate \texttt{str1:str2} together with a separator drawn from the three characters (\texttt{:,-}).  We then set \texttt{str1} to the resulting string, and \texttt{str2} to `'.  Thus, some tuples need to be correctly split before they can be cleaned to resolve functional dependency violations.

We consider two baseline libraries that each solve one type of error: \texttt{Split} only contains the string split operator, \texttt{String} only contains the string edit operators \texttt{ispell} and \texttt{edit\_dist\_match}.  \texttt{Comb} combines both libraries.  The quality function is the sum of the number of functional dependency violations, spelling errors, and empty strings.  We run \sys with 16 threads.

Figure \ref{exp10} shows that \texttt{Comb} takes longer than the baselines, but is capable of converging to a higher quality over all solutions.
We find that the asynchrony does not affect the sequential dependency and order of operations.
This is because of the tree-search, the operations that improve the quality score will be applied first and those that do not will be ignored.
These ignored operations may later become relevant in later rounds of the algorithm.  It is possible to have degenerate cases that mislead the pruning model, such as if {\it every} tuple must first be split before string edit fixes have any effect.  However, this is unlikely othewise.

{\it \stitle{Takeaways} \sys is designed to explore the plan space by leveraging the structure of data cleaning problems and out-performs generic blackbox parameter tuners.  Evidence suggests that \sys scales across cores, is robust to many forms of delays or redundancies, but is highly sensitive to slow quality evaluation.  Designing a system that adjusts to slow operators or quality evaluations is a promising direction for future work.}

\begin{figure}[h]
\centering
 \includegraphics[width=0.7\columnwidth]{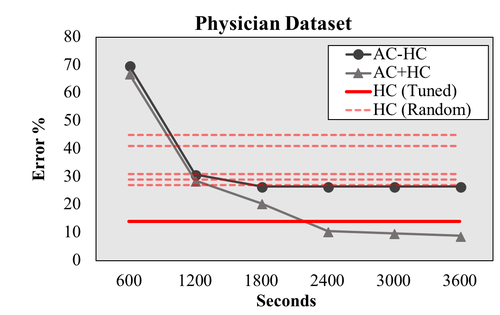}
 \includegraphics[width=0.7\columnwidth]{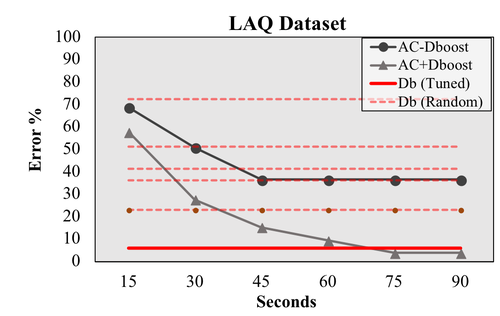}
 \caption{We compare \sys on the physician dataset and the air quality dataset against single standalone systems that address functional dependencies (Holoclean HC) and numerical errors (DBoost) respectively. \sys can support both types of errors and wrap around a variety of frameworks, and tune these frameworks.  Standalone system performance on 5 random parameters is shown as dashed lines.   \label{exp9}}
\end{figure}

\subsection{Comparison w/ Standalone Systems}
We now compare \sys with 2 standalone cleaning systems optimized for specific classes of  errors: HoloClean~\cite{rekatsinas2017holoclean} cleans functional dependency violations in the Physician data and dBoost~\cite{mariet2016outlier} detects numerical errors (we use last known good value as the replacement) in the LAQ data.  We compare \sys with the default library, the standalone system, and \sys with the standalone system wrapped as a cleaning operator. Note that \sys's quality function expresses both benchmarks, whereas each standalone system only expresses one of the two.

Figure \ref{exp9}a-b illustrates the results.
Even when a single data cleaning method can directly optimize the quality specification (i.e., integrity constraints), it might be beneficial to apply \sys to address the weak spots of the method. On the physician dataset, Holoclean (HC) achieves an accuracy of 86\% on its own, \sys without using Holoclean (AC-HC) achieves 73\%, and with using Holoclean (AC+HC) achieves 91\%. Similarly, on the air quality dataset, AC+DBoost achieves the best results and an even higher accuracy that DBoost on its own.
Furthermore, the standalone systems themselves are difficult to tune. 
Figure \ref{exp9}a-b plot the best version we found through manual parameter tuning (\red{solid lines}), as well as 5 runs with randomly sampled parameter values (\red{dashed lines}).  We find that the random parameters are highly unpredictable and often generate far worse results than either \sys variants.

{\it \stitle{Takeaways} \sys can model standalone systems as cleaning operators and improve the quality more than \sys or the standalone system on their own.   }

\section{Related Work}
Data cleaning is nearly as old as the relational model~\cite{codd1970relational}, and numerous research and  commercial systems have been proposed to improve data cleaning efficiency and accuracy (see~\cite{rahm2000data} for a survey).
The recent advances in scalable data cleaning~\cite{wang1999sample, DBLP:journals/debu/KrishnanWFGKM015, khayyat2015bigdansing, altowim2014progressive, he2016interactive, rekatsinas2017holoclean} has revealed {\it human-time}---finding and understanding errors, formulating desired characteristics of the data, writing and debugging the cleaning pipeline, and basic software engineering---as a dominant bottleneck in the entire data cleaning process~\cite{krishnan2016hilda}.  
\sys aims to address this bottleneck by using the quality function and conditional assignment API as a flexible and expressive declarative interface to separate high level cleaning goals from how the goals are achieved.  

\stitle{Machine Learning in Data Cleaning} Machine learning has been widely used to improve the efficiency and/or reliability of data cleaning~\cite{DBLP:journals/pvldb/YakoutENOI11,yakout2013don,gokhale2014corleone}.
It is commonly used to predict an appropriate replacement attribute value for dirty records~\cite{yakout2013don}.
Increasingly, it is used in combination with crowd-sourcing to extrapolate patterns from smaller manually-cleaned samples~\cite{gokhale2014corleone,DBLP:journals/pvldb/YakoutENOI11} and to improve reliability of the automatic repairs~\cite{DBLP:journals/pvldb/YakoutENOI11}.
Concepts such as active learning can be leveraged to learn an accurate model with a minimal number of examples~\cite{DBLP:journals/pvldb/MozafariSFJM14}.

For example, Yakout et al. train a model that evaluates the likelihood of a proposed replacement value \cite{yakout2013don}.
Another application of machine learning is value imputation, where a missing value is predicted based on those records without missing values.
Machine learning is also increasingly applied to make automated repairs more reliable with human validation \cite{DBLP:journals/pvldb/YakoutENOI11}.
 Human input is often expensive and impractical to apply to entire large datasets.
Machine learning can extrapolate rules from a small set of examples cleaned by a human (or humans) to uncleaned data \cite{gokhale2014corleone, DBLP:journals/pvldb/YakoutENOI11}.
This approach can be coupled with active learning \cite{DBLP:journals/pvldb/MozafariSFJM14} to learn an accurate model with the fewest possible number of examples.
Holoclean~\cite{rekatsinas2017holoclean} leverages machine learning to validate repairs with a probabilistic graphical model.
 \sys uses machine learning in the synthesis process to prune search branches.
 We see \sys as complimentary to these techniques: as increasingly sophisticated cleaners have more opaque parameters, meta algorithms such as \sys can help tune and compose them.

\stitle{Application-Aware Cleaning}  Semantics about the downstream application can inform ways to clean the dataset ``just enough'' for the application.  
A large body of literature addresses relational queries over databases with errors by focusing on specific classes of queries~\cite{altwaijry2015query}, leveraging constraints over the input relation~\cite{2011Bertossi}, integration with crowd-sourcing~\cite{DBLP:conf/sigmod/BergmanMNT15}.   Recent work such as ActiveClean~\cite{DBLP:journals/pvldb/KrishnanWWFG16} extend this work to downstream machine learning applications, while Scorpion~\cite{DBLP:journals/pvldb/0002M13} uses the visualization-specified errors to search for approximate deletion transformations.   In this context, \sys can embed application-specific cleaning objectives within the quality function.  For instance, our the london air quality benchmark simply embeds an autoregression model into the the quality function. 
Recent work on quantifying incompleteness in data quality metrics~\cite{chung2016data} suggests that the flexibilty to embed new quality measures is of practical value.

\stitle{Generating Cleaning Programs} A composable data cleaning language is the building block for systems like \sys that generate cleaning pipelines.   Languages for data transformations have been well-studied, and include seminal works by Raman and Hellerstein~\cite{raman2001potter} for schema transformations and Galhardas et al.~\cite{DBLP:conf/vldb/GalhardasFSSS01} for declarative data cleaning. These ideas were later extended in the Wisteria project~\cite{DBLP:journals/pvldb/HaasKWF015} to parameterize the transformations to allow for learning and crowdsourcing.   Wrangler~\cite{wrangler} and Foofah~\cite{jin2017foofah} are text extraction and transformation systems that similarly formulate their problems as search over a language of text transformations, and develop manual pruning heuristics to reduce the search space. We do not intend for \sys to be applied to schema transformation problems and design \sys around existing patterns observed in data cleaning.
We defer the study of a broader programming-by-example data cleaning suite to future work.

\section{Conclusion and Future Work}
The research community has developed increasingly sophisticated data cleaning methods~\cite{dc, rekatsinas2017holoclean, DBLP:journals/pvldb/KrishnanWWFG16, DBLP:conf/sigmod/ChuIKW16, mudgal2018deep, doan2018toward}.
The burden on the analyst is gradually shifting away from the design of hand-written data cleaning scripts, to building and tuning complex pipelines of automated data cleaning libraries.
The main insight of this paper is that tuning pipelines of data cleaning operations is very different than tuning pipelines for machine learning.

Rather than treat each pipeline component as a black-box transformation of the relation, \sys canonicalizes their {\it  repairs} as conditional assignment operations.   Given a library of  cleaning operators, their outputs contribute to a pool of conditional assignments. This defines a well-posed search space, namely, the set of all pipelines composed of conditional assignments.  %
  
Although our results suggest that leveraging advances in planning and optimization can solve a range of data cleaning benchmarks, they are counter-intuitive because of greedy nature of the system and its enormous search space.  This raises a number of questions about future opportunities in data cleaning.  Why does a greedy search achieve strong results on widely-used cleaning benchmarks? Are the benchmarks too simple or are cleaning problems simply highly structured?  We hope to understand the fundamental reasons for when and why search-based approaches should perform well.

In addition, we are excited to extend \sys towards a more flexible, visual, and interactive cleaning process.  We plan to integrate \sys with a data visualization system~\cite{Wu2017CombiningDA} so users can visually manipulate data visualizations that are translated into quality functions.  This will also require work to characterize failure modes and provide high-level tools to debug such cases.

{
\bibliographystyle{abbrv}
\bibliography{main} 
}

\end{document}